\documentclass[a4paper]{article}

\usepackage{INTERSPEECH2019}

\usepackage{scalerel}% for \scaleto
\usepackage[justification=centering]{caption} % to center the caption
\usepackage{multirow, makecell} % for multirows in table
\usepackage{amsmath}

\usepackage[labelformat=simple]{subcaption}

\newcommand{\mcep} {\textit{mcp}}

\newcommand{\sig}[2] {\textit{sig}^{(#1)}_{#2}}
\newcommand{\res}[1] {\textit{res}^{(#1)}}
\newcommand{\mcp}[2] {\textit{mcp}^{(#1)}_{#2}}
\newcommand{\env}[1] {\textit{env}_{#1}}

\title{Generalization of Spectrum Differential based Direct Waveform Modification for Voice Conversion}
\name{ \begin{tabular}{c}
		Wen-Chin Huang$^1$,
		Yi-Chiao Wu$^1$,
		Kazuhiro Kobayashi$^1$,
		Yu-Huai Peng$^2$,
		Hsin-Te Hwang$^2$,\\
		Patrick Lumban Tobing$^1$,
		Yu Tsao$^2$,
		Hsin-Min Wang$^2$
		Tomoki Toda$^1$,
	\end{tabular}
}
\address{$^{1}$ Nagoya University, Japan\\
		 $^{2}$ Academia Sinica, Taiwan
}

\email{wen.chinhuang@g.sp.m.is.nagoya-u.ac.jp}

\begin{document}

\maketitle
\begin{abstract}
	We present a modification to the spectrum differential based direct waveform modification for voice conversion (DIFFVC) so that it can be directly applied as a waveform generation module to voice conversion models. The recently proposed DIFFVC avoids the use of a vocoder, meanwhile preserves rich spectral details hence capable of generating high quality converted voice. To apply the DIFFVC framework, a model that can estimate the spectral differential from the F0 transformed input speech needs to be trained beforehand. This requirement imposes several constraints, including a limitation on the estimation model to parallel training and the need of extra training on each conversion pair, which make DIFFVC inflexible. Based on the above motivations, we propose a new DIFFVC framework based on an F0 transformation in the residual domain. By performing inverse filtering on the input signal followed by synthesis filtering on the F0 transformed residual signal using the converted spectral features directly, the spectral conversion model does not need to be retrained or capable of predicting the spectral differential. We describe several details that need to be taken care of under this modification, and by applying our proposed method to a non-parallel, variational autoencoder (VAE)-based spectral conversion model, we demonstrate that this framework can be generalized to any spectral conversion model, and experimental evaluations show that it can outperform a baseline framework whose waveform generation process is carried out by a vocoder.	
	
\end{abstract}

\section{Introduction}

Voice conversion (VC) aims to convert the speech from a source to that of a target without changing the linguistic content. Numerous approaches have been proposed, such as Gaussian mixture model (GMM)-based methods \cite{VC,GMM-VC}, deep neural network (DNN)-based methods \cite{ANN-VC, layerwise-VC}, and exemplar-based methods \cite{exemplar-noisy-VC,exemplar-residual-VC,LLE-VC}. While most VC researchers have focused on the conversion of spectral features, the waveform generation process, in fact, plays an important role in a VC system. Conventional VC frameworks employ parametric vocoders \cite{LPC, STRAIGHT, WORLD} as their synthesis module, which impose many overly simplified assumptions that discard the phase information and result in unnatural excitation signals, and thus cause a significant degradation in the quality of the converted speech.

In recent years, there are two mainstreams that try to improve the waveform generation module. One direction is to develop neural vocoders \cite{SDWN, SIWN, WaveRNN, SampleRNN, FFTNet, LPCNet, Parallel-WN, WaveGlow, NSF}, which are capable of reconstructing the phase and excitation information, and thus generate extremely natural sounding speech. State-of-the-art VC systems have shown remarkable results by combining such neural waveform generation process with conversion models such as GMMs \cite{VCWN}, or other methods based on recent DNN methods such as bidirectional long short term memory \cite{Mel-WNV-VC}, autoencoders \cite{AUTOVC}, sequence-to-sequence learning \cite{seq2seq-vc-iFLYTEK} or generative adversarial networks \cite{GAN-WNV-VC}. Nonetheless, neural vocoders usually require a large amount of training data and are computationally expensive.

\begin{figure}[t]
  \centering
  \includegraphics[width=0.48\textwidth]{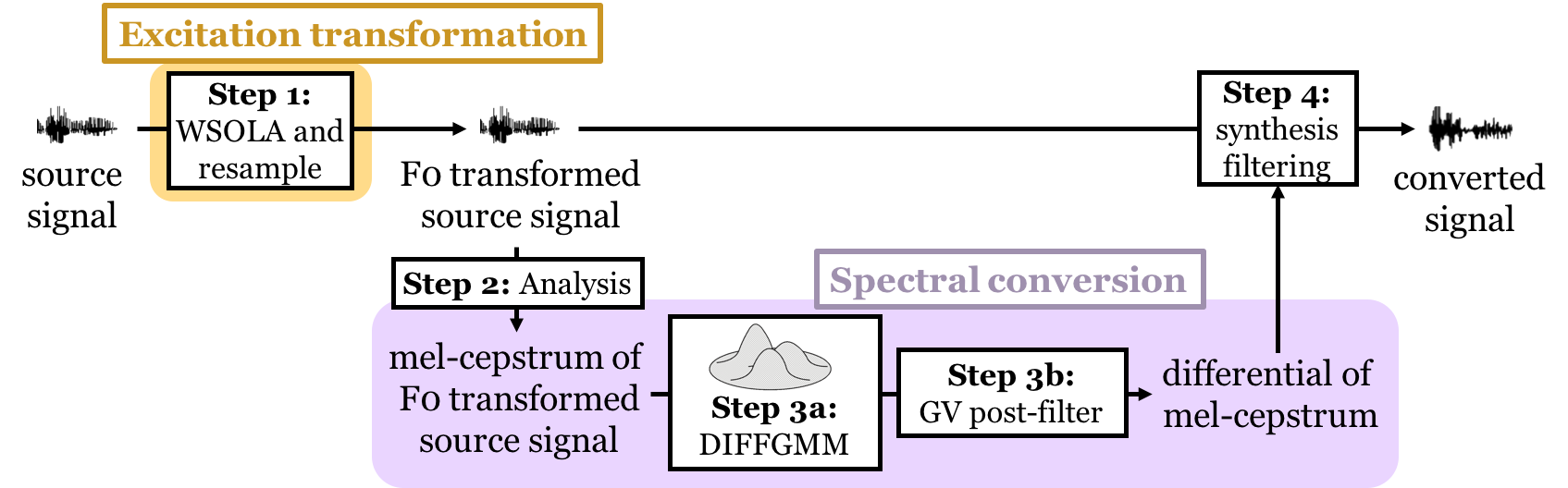}
  \centering
  \captionof{figure}{The spectrum differential based direct waveform modification for vocoder free voice conversion, where the spectrum differential is estimated using a Gaussian mixture model (GMM). Such a GMM model is termed \textit{DIFFGMM}.}
  \label{fig:diffgmm}
\end{figure}

Another approach which is simple, efficient yet free from any extra training data is the direct waveform modification framework based on spectrum differential, which was first applied to VC in \cite{NU-VCC2016} (DIFFVC). The DIFFVC framework first estimates a sequence of spectral differential from the F0 transformed signal with a trained differential GMM (DIFFGMM), The converted speech is then obtained by directly filtering the F0 transformed speech with the spectral differential. In the Voice Conversion Challenge (VCC) 2016 \cite{vcc2016}, a predecessor version \cite{NU-VCC2016} was evaluated as one of the best systems achieving the best conversion accuracy on speaker identity and high speech quality. It was then adopted as the baseline system of the VCC2018, and achieved the second highest sound quality for the same-gender speaker conversion pairs \cite{sprocket}.

Despite the success of DIFFVC, one flaw is its inflexibility. First, in order to train a model that can estimate the spectral differential, parallel data is needed. This makes application of DIFFVC to non-parallel VC methods infeasible. Moreover, the differential estimation model in the state-of-the-art DIFFVC framework needs to be trained using the features extracted from the F0 transformed signal. In other words, for an arbitrary VC model that we wish to apply DIFFVC to, it requires retraining the model for each source-target conversion pair using the corresponding F0 transformed features. This limitation significantly increases the inflexibility of the framework.

To this end, our goal is to develop a flexible method which is free from parallel data or any extra training process so that one can directly apply the vocoder-free DIFFVC framework to an arbitrary VC model. Our proposed method is based on the residual signal modification based F0 transformation implementation, which is similar to the method proposed in \cite{DIFFVC-F0trans-implement}. Our contributions are:
	\begin{itemize}
		\item We generalize the DIFFVC framework such that it is applicable to any VC model as long as it is able to perform spectral conversion from the normal source speech, regardless of how it was trained. In other words, the VC model need not to estimate the spectrum differential nor use the F0 transformed features as input, thus free from parallel data or extra retraining procedure.
		\item We propose several techniques to address some details that need to be taken care of when adopting the residual signal modification based F0 transformation, such as power compensation, collapsed speech detection and feature substitution.
	\end{itemize}

As a proof of concept, we show the effectiveness of our proposed method by applying it to a variational autoencoder based VC model, which we will refer to as VAE-VC \cite{VAE-VC}. It was trained with non-parallel data, and takes normal, ``un-F0-transformed'' as input during conversion. Through subjective evaluations, we show that our proposed framework can bring a significant performance boost in some conversion gender pairs, compared to a baseline system which employs a conventional parametric vocoder, WORLD \cite{WORLD}, in terms of speech naturalness and conversion accuracy. Note that our goal is to demonstrate the ability of our method to generalize to any VC model, so it is not restricted to VAE-VC. 

\begin{figure*}[t]
  \centering
  \includegraphics[width=\textwidth]{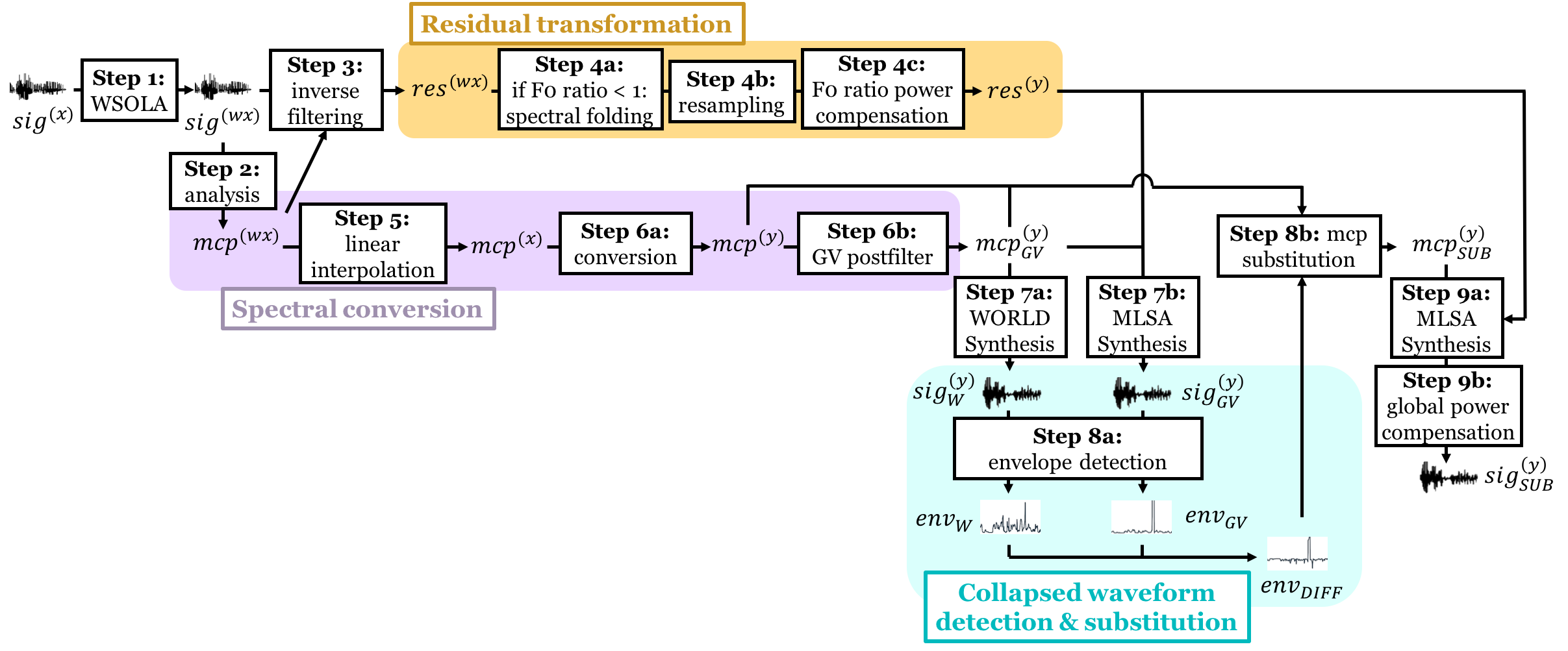}
  \centering
  \captionof{figure}{The proposed direct waveform modification framework, where the F0 transformation is realized in the residual domain. \textit{sig}, \textit{res}, \textit{mcp}, and \textit{env} represent the waveform signal, residual signal, mel-cepsturm, and envelope signal, respectively.}
  \label{fig:diffres}
\end{figure*}

\section{Spectrum Differential based Direct Waveform Modification for Voice Conversion (DIFFVC)}
\label{sec:DIFFVC}

\subsection{DIFFVC based on DIFFGMM}

DIFFVC is a conversion framework (not restricted to VC but also other applications like singing VC) that does not employ a parametric vocoder as the waveform generation module \cite{NU-VCC2016, sprocket, DIFFVC-F0trans-implement, DIFFVC-F0trans-tech}. In this section we describe the DIFFVC framework in \cite{DIFFVC-F0trans-tech}. In the offline stage, instead of training a VC model capable of mapping source features to target features, DIFFVC requires to model the spectral differential between the source and target. The differential is obtained by subtracting the target feature with the source feature. To realize this, a joint density GMM \cite{GMM-VC} was first trained using the joint vectors of source and target features in the parallel dataset. Then, the DIFFGMM can be analytically derived by transforming the trained model parameters. Note that to train such a model, a \textit{parallel} dataset is required, which is a fundamental limitation of this method.
%Here the mel cepstrum coefficient (\mcep) is adopted as the spectral feature. 

The conversion process is depicted in Figure~\ref{fig:diffgmm}. First, an F0 transformation process is performed to obtain the F0 transformed waveform. Then, the \mcep\ differential is estimated via maximum likelihood speech parameter estimation \cite{GMM-VC} based on the DIFFGMM, optionally enhanced by a global variance (GV) postfilter \cite{fastGV}. Finally, the converted speech is obtained by directly filtering the F0 transformed signal with the differential.

\subsection{F0 transformation techniques}

In this subsection, we describe two F0 transformation techniques proposed in \cite{DIFFVC-F0trans-tech}. Note that in both techniques, we assume the F0 transformation ratio is time-invariant and set it to a constant value, which is calculated using the training data of the source and target speakers.

\subsubsection{F0 transformation by residual signal modification}
\label{sssec:F0transform-res}

In this method, the F0 transformation is performed by modifying the residual signal. According to the source-filter model, given a series of waveform signals $S(z)$ and the spectral envelope $H(z)$, the ideal excitation signal (or \textit{residual}) $E(z)$ can first be computed by filtering $S(z)$ with the inverse filter $H(z)$:
\begin{equation}
	E(z)=\frac{1}{H(z)}S(z),
\end{equation}
where the spectral features extracted from $S(z)$ are employed as coefficients of the time-invariant filter $H(z)$. Then, a WSOLA \cite{wsola} and resampling process can be performed on the residual signal in order to transform F0. Specifically, the residual signal is shrunk then up-sampled if the F0 transformation ratio is smaller than 1 and, conversely, expanded then down-sampled if the ratio larger than 1. Finally, the F0 transformed speech is restored by filtering the modified residual signal using the same $H(z)$.

As discussed in \cite{DIFFVC-F0trans-implement}, when the ratio is smaller than 1, the high frequency components of the transformed residual signal vanish, thus need to be reconstructed. \cite{DIFFVC-F0trans-implement} proposed to reconstruct them by adding a high-pass filtered noise excitation signal to the F0 transformed residual signal, as they claimed that the high frequency components of a speech signal tend to be less periodic and well modeled with noise components. 

This technique has several flaws. First, the WSOLA process needs to be applied to the residual signal, which presents a challenge for WSOLA since it was originally designed for normal speech. In addition, the generation of the high-pass filtered noise excitation signal is time-consuming, and sometimes causes discontinuities or asynchronous conditions with low frequency residual components. Finally, by using this technique, it is assumed that the residual spectrum is perfect, i.e., its spectral tilt is totally flat, which is not in practice due to imperfect inverse filtering. As a result, this method does not work well as reported in \cite{DIFFVC-F0trans-implement}. Our proposed method addresses these issues, as described in Section~\ref{ssec:DIFFRES-high-freq}.

\subsubsection{F0 transformation by direct waveform modification}
\label{sssec:F0transform-wav}

In this method, rather than performing the complicated chain process (\textit{inverse-transform-synthesis}) described in Section~\ref{sssec:F0transform-res}, the F0 transformation process is directly applied to the source waveform. 
%Here, the same high frequency component reconstruction process described in Section~\ref{sssec:F0transform-res} is needed when the F0 transformation ratio is less than 1.
The advantages of this process are that first, no approximation errors caused by the processes like inverse filtering are introduced. It can be therefore expected that a high-quality transformed signal can be obtained. Moreover, since this method is simpler, it is more likely to be embedded to real-time VC systems.

However, this direct waveform modification causes a frequency warping issue, so the DIFFGMM needs to be trained with features extracted from the F0 transformed and natural target speech. In other words, a separate DIFFGMM for each F0 transformation ratio needs to be trained because the spectral envelope of the F0 transformed signal depends on the F0 transformation ratio. 
%Note that fortunately, although the high frequency components are still generated with aliasing in this method, the resulting problematic spectral envelope is modeled by the DIFFGMM, thus does not cause quality degradation during conversion.

\section{Proposed Method based on Residual Transformation}

Our goal is to extend the vocoder-free DIFFVC framework to any arbitrary VC model, which only knows how to convert normal source features to target features. To impose as few constraints as possible, we only demand the VC model to estimate the converted spectral features given source spectral features extracted from a normal source speech, regardless of whether a parallel training dataset is available.
%A desired framework should be flexible enough to be free from parallel data or any extra training process.
In this section, we describe our proposed method, as depicted in Figure~\ref{fig:diffres}. 

\subsection{Interpretation of F0 transformation based on residual transformation}
\label{ssec:DIFFRES-F0-trans}

The key idea of our proposed method resorts back to the motivation of using the spectrum differential, which is based on the source-filter model. To obtain the ideal time-domain, spectrally converted speech signal $y[n]$, consider the following decomposition:
\begin{align}
	y[n] & = h_{y|x}[n] \ast x[n] \label{eq:diff-direct-est} \\
	&\approx h_y[n] \ast h_x[n]^{-1} \ast x[n] \label{eq:diff-two-step} ,
\end{align}
where $x[n]$ is the input time-domain source speech signal. Eq.~\eqref{eq:diff-direct-est} is realized by filtering $x[n]$ with a time-variant filter $h_{y|x}[n]$, whose coefficients (the spectrum differential) are estimated by a VC model. This is exactly the DIFFVC method in Section~\ref{sec:DIFFVC}. Then, we can derive an equivalent form, as shown in Eq.~\eqref{eq:diff-two-step}. This equation can be realized by first performing inverse filtering with coefficients $h_x[n]$ to $x[n]$ (step 3), followed by synthesis filtering with coefficients $h_y[n]$ (steps 7b and 9a). It is thereby obvious that this method can be flexibly applied whenever a VC model is available, as we expected.

Now it is crucial to decide where to apply the F0 transformation process (WSOLA + resampling), which we will denote as $f(\cdot)$. In the implementation of \cite{DIFFVC-F0trans-tech}, to minimize the effect of frequency wrapping caused by resampling, the F0 transformation process is applied to the residual domain. We formulate this by modifying Eq.~\eqref{eq:diff-two-step}:
\begin{equation}
	y[n] \approx h_y[n] \ast f(h_x[n]^{-1} \ast x[n]).
\end{equation}
As reported in \cite{DIFFVC-F0trans-tech}, the residual based F0 transformation is inferior to the waveform based process described in Section~\ref{sssec:F0transform-wav} in terms of performance. Nonetheless, we would like to emphasize that the vast limitations of the waveform based transformation greatly conflict with our initial purpose to generalize the DIFFVC framework. In contrast, although somehow problematic, the residual based process is superior in terms of flexibility, which is the main reason we adopt it since it is a suitable choice for our goal. To overcome the various problems of the residual based transformation described in Section~\ref{sssec:F0transform-res}, in the rest of this section, we propose several techniques to tackle with the issues.

\subsection{An alternative of high frequency component reconstruction}
\label{ssec:DIFFRES-high-freq}

We propose a different approach to reconstruct the high frequency components of the transformed residual when the F0 ratio is smaller than 1. After performing WSOLA, we insert zeroes between every two samples in $\res{wx}$ , which is called upsampling by zero stuffing or spectral folding \cite{high-freq-gen}. This process doubles the cutoff frequency and generates a symmetric spectral envelope with respect to the original cutoff frequency. Such technique has been applied to various speech processing fields, such as bandwidth extension \cite{spectral-folding-bandwidth-extension}. Then, the resampling process discards the spectral envelope with frequency larger than $\textit{cutoff frequency}/\textit{F0 ratio}$. As a result, the missing frequency components are generated by copying a mirrored part of that of $\res{wx}$. Since the reconstructed component is merely a reversed copy, it is expected to be more continuous. Another advantage of this approach is that it is much simpler to implement. We formulate the aforementioned changes into the following decomposition:
\begin{align}
	x'[n] &= f_w(x[n])\\
	y[n] &= h_y[n] \ast f_r(h_{x'}[n]^{-1} \ast x'[n]),
\end{align}
where we decompose $f(\cdot)$ into two functions $f_w(\cdot)$ and $f_r(\cdot)$, the former being the WSOLA-based duration conversion function (step 1) and the latter being the resampling function (step 4b), preceded by the spectral folding technique (step4a) if needed. One trick used here is that, since \mcep\ extraction (step2) is time-consuming, we try to utilize $\mcp{wx}{}$ as much as possible, by using it as the input feature to the conversion process (step 6a). This can be accomplished by applying linear interpolation to $\mcp{wx}{}$ to restore the original time length (step 5).

\begin{figure}[t]
  \centering
  \includegraphics[width=0.48\textwidth]{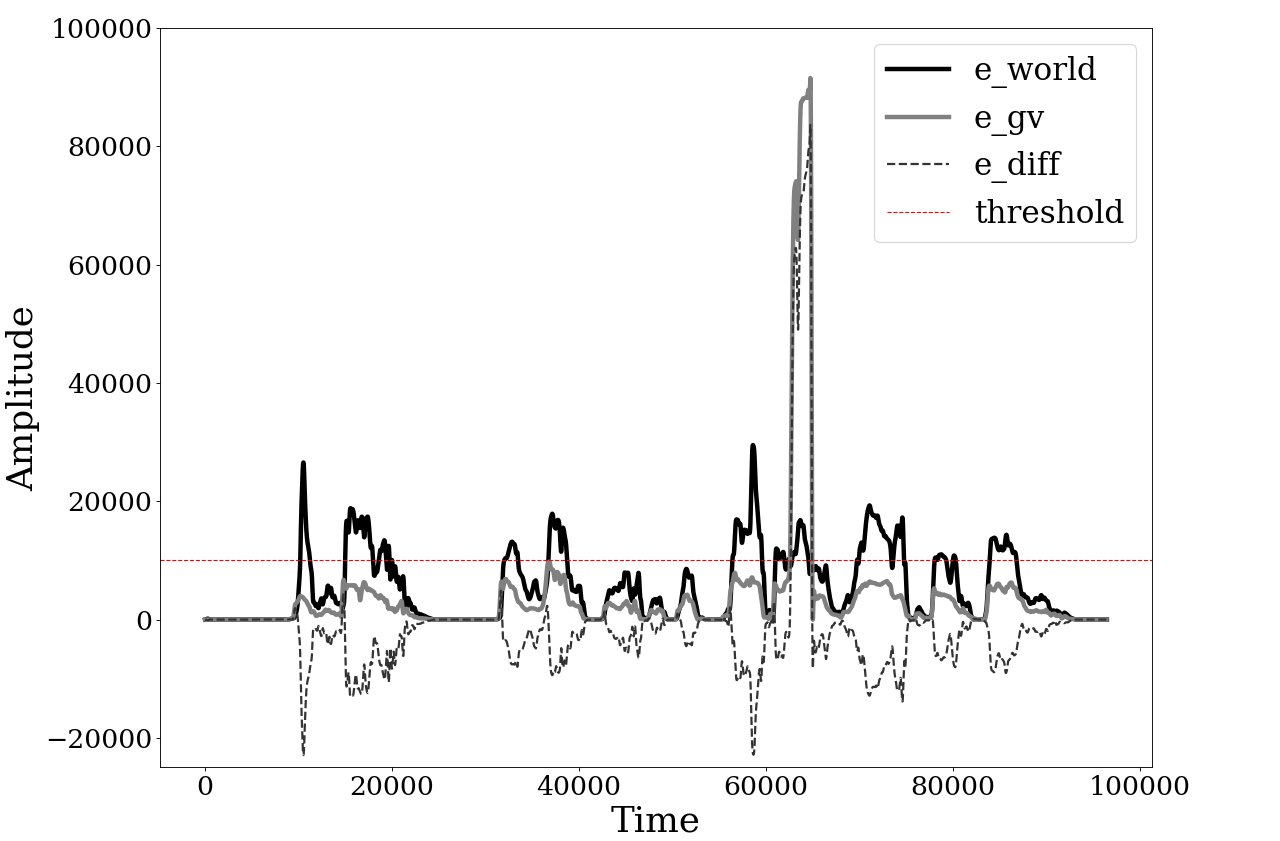}
  \centering
  \captionof{figure}{An illustration of the calculated envelopes $\env{W}$ (\textnormal{e\_world}), $\env{GV}$ (\textnormal{e\_gv}) and $\env{DIFF}$ (\textnormal{e\_diff}), which are from the converted sample SF1-TF1-30006. The red dashed line denotes the threshold, which is set to be 10000 here.}
  \label{fig:envelopes}
\end{figure}

\subsection{Collapsed waveform detection and feature substitution}

In our initial experiments, we often observed collapsed waveform segments in $\sig{y}{GV}$. This is a combined result of the frequency axis warping effect caused by resampling, as pointed out in \cite{DIFFVC-F0trans-implement, DIFFVC-F0trans-tech}, and the imperfect residual signal obtained from filtering. During conversion, a GV postfilter could further amplify this effect, leading to greater instability.

As a remedy, we replace the postfiltered features $\mcp{y}{GV}$ that cause collapsed segments with the corresponding ones without postfiltering $\mcp{y}{}$. To detect collapsed waveform intervals, we adopt a modified signal envelope extraction method proposed in \cite{LPCWNV}. Specifically, the waveform signal is first passed through a Hilbert transform, which is often used to extract the signal envelope. Then, it is divided into non-overlapping slots with a fixed window, and the maximum value of each slot is used to replace every value in that slot. Finally, a low-pass filter is used to smoothen the curve. Figure~\ref{fig:envelopes} illustrates the calculated envelopes.

The complete detection and substitution procedures can be summarized as follows. First, we use the WORLD vocoder to generate $\sig{y}{W}$ from $\mcp{y}{GV}$ as reference (step 7a). Then, the above process is applied to both $\sig{y}{W}$ and $\sig{y}{GV}$ to obtain the corresponding envelopes $\env{W}$ and $\env{GV}$ (step 8a). The collapsed interval is then detected by setting an empirically set threshold to the difference between the envelopes. Finally, feature substitution is performed to obtain $\mcp{y}{SUB}$ (step 8b), which can be used as the coefficients of the final synthesis filtering process.

\subsection{Power compensation}

It is observed that the power of the signal changes after resampling. We perform two power compensation processes. In step 4c, the signal power is compensated according to the F0 transformation ratio:
\begin{equation}
	\res{y} = \res{y} \cdot \sqrt{\frac{1}{\textit{F0 ratio}}},
\end{equation}
while in step 9b, the signal power is normalized to be the same as the input speech:
\begin{equation}
	\sig{y}{SUB} = \sig{y}{SUB} \cdot \sqrt{\frac{\sum_n \sig{x}{}[n] ^2}{\sum_n \sig{y}{SUB}[n] ^2}}.
\end{equation}

\section{Experimental Evaluation}
\label{sec:exp}

\subsection{Experimental settings}
 
We evaluated our proposed methods on the SPOKE task of Voice Conversion Challenge 2018 (VCC2018) \cite{vcc2018}, which included recordings of professional US English speakers with a sampling rate of 22050 Hz. The dataset consisted of 81/35 utterances per speaker for training/testing sets, respectively. We used the first 70 utterances of the training set of all speakers for training, the remaining 11 for validation. A total of four source speakers and four target speakers formed 16 conversion pairs for evaluation. The WORLD vocoder \cite{WORLD} was adopted to extract acoustic features including 513-dimensional spectral envelopes (SPs), 513-dimensional aperiodicity signals (APs) and F0. 35-dimensional \mcep\ were further extracted from the SPs. 

The conversion model adopted in this work is a fully convolutional cross domain VAE (FCN-CDVAE) \cite{F0-FCN-CDVAE}, which achieves non-parallel VC from an unsupervised factorization of spectral speech frames via autoencoding. In the model, conversion was carried out by encoding speaker independent latent vectors and decoding with the desired target speaker representation\footnote{Official implementation: https://github.com/unilight/cdvae-vc}. The digital filtering were mainly implemented with the open source sprocket software \cite{sprocket}. We thereby choose \mcep\ as the spectral feature, and the mel log spectrum approximation (MLSA) filter \cite{MLSA} for both inverse and synthesis filtering. To generate $\bm{y_{\scaleto{W}{3pt}}}$, following the original VAE-VC \cite{VAE-VC}, the APs and energy of \mcep\ were kept unmodified, and the F0 was converted using a linear mean-variance transformation in the log domain. For feature substitution, the threshold was set to 10000.

\begin{table}[t]
	\centering
	\captionsetup{justification=centering}
	\caption{Subjective evaluation results of the FCN-CDVAE based VC method \cite{F0-FCN-CDVAE} with a waveform generation process by the WORLD vocoder or the proposed method. \\Here M and F denotes male and female, respectively.}
	\centering
	\begin{tabular}{ l c c c }
		\toprule	
		 & Proposed & WORLD & p-value\\
		\midrule
		\textbf{naturalness} & 46.9\% & \textbf{53.1\%} & 0.114 \\
		\;\;M-M & 30.6\% & \textbf{69.4\%} & $< 0.001$ \\
		\;\;M-F & 16.9\% & \textbf{83.1\%} & $< 0.001$ \\
		\;\;F-M & \textbf{63.8\%} & 36.2\% & $< 0.001$ \\
		\;\;F-F & \textbf{76.3\%} & 23.7\% & $< 0.001$ \\
		\midrule
		\textbf{similarity} & \textbf{63.8}\% & 36.2\% & $< 0.001$ \\
		\;\;M-M & \textbf{56.3\%} & 33.7\% & 0.113 \\
		\;\;M-F & 36.2\% & \textbf{63.8\%} & $< 0.001$ \\
		\;\;F-M & \textbf{77.5\%} & 22.5\% & $< 0.001$ \\
		\;\;F-F & \textbf{85.0\%} & 15.0\% & $< 0.001$ \\
	\end{tabular}
	\label{tab:preference}
\end{table}

\subsection{Subjective evaluations}
\label{ssec:exp}

The goal of the proposed framework is to generalize the DIFFVC framework to an arbitrary, specifically non-parallel VC model, in order to improve the naturalness of the converted speech by avoiding the vocoding process. To confirm the effectiveness of our proposed framework, we conducted two preference tests to compare the naturalness as well as the conversion similarity of the waveforms generated using a conventional high-quality vocoder, WORLD, or the proposed framework. Note that here a GV postfilter was used to enhance the output of WORLD. Speech samples can be found in: https://unilight.github.io/Publication-Demos/publications/ssw10/index.html

In the naturalness test, each participant was demanded to choose the one with more natural voice among two converted utterances generated by the two methods for the same sentence (content) in random order. In the conversion similarity test, a natural speech sample of the target speaker was first presented as a reference. Then, each participant decided which of the two converted utterances generated by the two methods for the same sentence was more similar to the reference speech in terms of speaker identity. We recruited 10 non native English speakers for evaluation.

%\begin{figure*}[t]
%\begin{minipage}[b]{0.48\textwidth}
%  \centering
%  \includegraphics[width=\textwidth]{scatter_naturalness.png}
%  \captionof{figure}{Scatter plot of the relationship between F0 transformation ratio and naturalness preference.}
%  \label{fig:scatter-naturalness}
%\end{minipage}
%~
%\begin{minipage}[b]{0.48\textwidth}
%  \centering
%  \includegraphics[width=\textwidth]{scatter_similarity.png}
%  \captionof{figure}{Scatter plot of the relationship between F0 transformation ratio and similarity preference.}
%  \label{fig:scatter-similarity}
%\end{minipage}
%\end{figure*}

Table~\ref{tab:preference} shows the subjective evaluation results. The p-values are calculated with two-tailed t-tests. From the table, it is clear that the proposed method outperformed the use of WORLD as the former achieved comparable speech naturalness but significantly better conversion similarity in terms of overall score. More specifically, subjects had very high preference on our method for speech naturalness when converting from a female source speaker, while being inferior with a male source speaker. On the other hand, our method achieved significantly higher conversion similarity over WORLD in all but the M-F pair. This result implies that by avoiding the use of vocoders, the output speech can be made free from potential naturalness degradation, such as the error-prone phase estimation and buzzy noises. As a result, rich spectral details could be preserved in order to achieve high conversion accuracy. 

As reported in \cite{NU-VCC2016}, in terms of speech naturalness, the combination of a GMM-based parallel VC model with the DIFFVC method is usually superior to those employing conventional vocoders \cite{GMM-VC} for intra-gender speaker pairs, while there is no significant difference for the conversion similarity. This trend was, however, not observed when applying our method to the VAE-based VC model. We first attribute the difference in conversion accuracy to the fundamentally different F0 transformation process adopted. Since we performed this process in the residual domain, it is possible that the spectral envelope preserved included less speaker-dependent or gender-dependent features, therefore increasing the conversion accuracy. As for naturalness, we suggest several possible sources of degradation, as we will discuss in the next section.

%\begin{figure}[t]
%	\centering
%	\captionsetup{type=figure}
%		
%	\begin{subfigure}[b]{0.23\textwidth}
%		\centering
%		\includegraphics[width=\textwidth]{SM1-30001.png} 
%		\caption{SM1-30001
%		\label{fig:SM1_30001_sgram}}
%	\end{subfigure}
%	\hfill
%	\begin{subfigure}[b]{0.23\textwidth}
%		\centering
%		\includegraphics[width=\textwidth]{yh-SM1-TF1-30001-sub.png} 
%		\caption{SM1-TF1-30001
%		\label{fig:SM1_TF1_30001_sgram}}
%	\end{subfigure}
%	
%	%	\vspace{0.7cm}
%	
%	\begin{subfigure}[b]{0.23\textwidth}
%		\centering
%		\includegraphics[width=\textwidth]{SF1-30001.png} 
%		\caption{SF1-30001
%		\label{fig:SF1_30001_sgram}}
%	\end{subfigure}
%	\hfill
%	\begin{subfigure}[b]{0.23\textwidth}
%		\centering
%		\includegraphics[width=\textwidth]{yh-SF1-TF1-30001-sub.png} 
%		\caption{SF1-TF1-30001
%		\label{fig:SF1_TF1_30001_sgram}}
%	\end{subfigure}
%	
%	\centering
%	\caption{Spectrograms of the converted speeches using our proposed method. 
%		\label{fig:sgrams}}
%\end{figure}

\begin{figure}[t]
  \centering
  \includegraphics[width=0.48\textwidth]{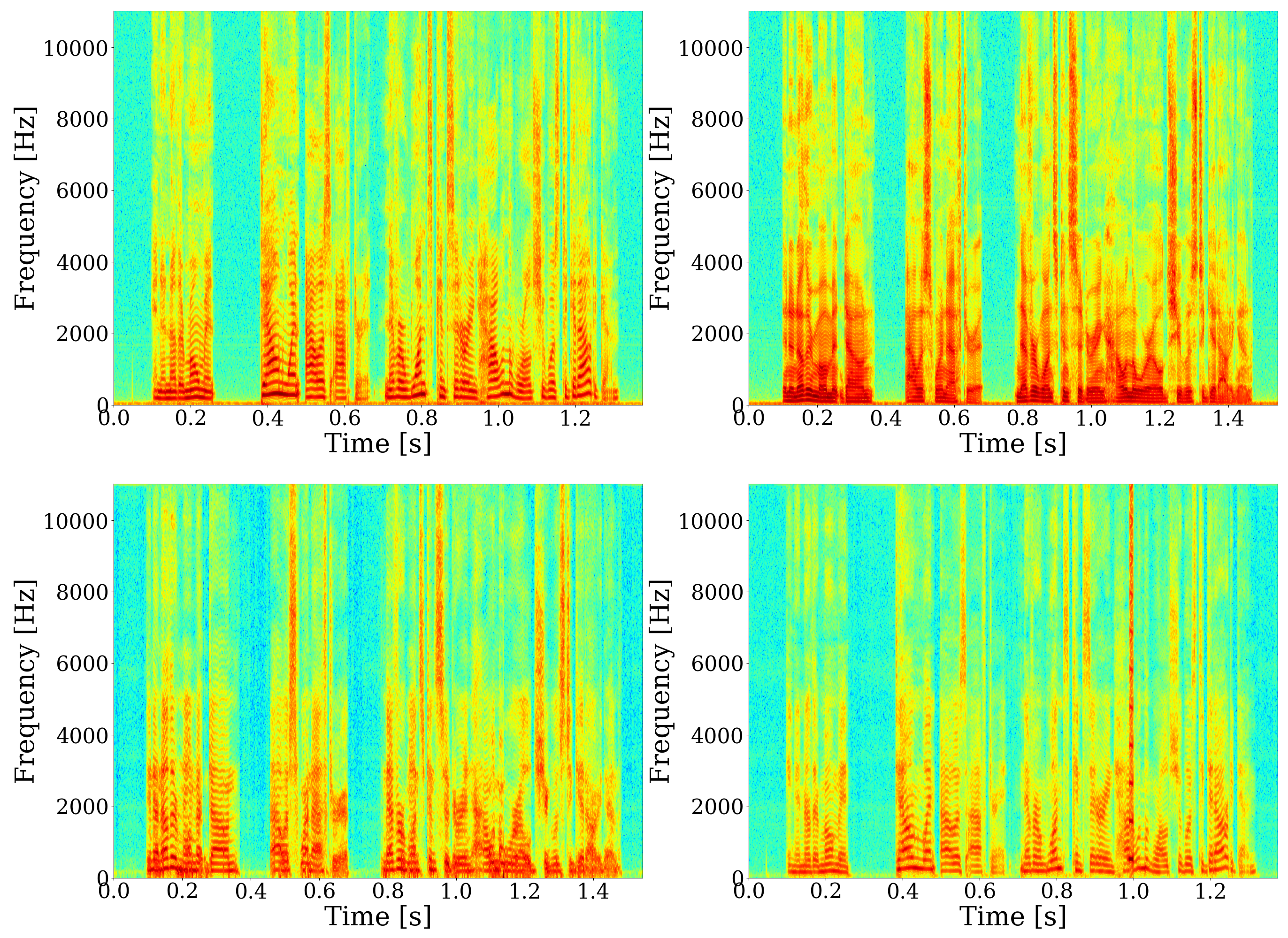}
  \centering
  \caption{The top, bottom rows are the spectrograms of the source, converted speeches using our method, respectively. \\ 
  		\textit{Top left}: SF1-30001. \textit{Top right}: SM1-30001. \\
  		\textit{Bottom left}: SF1-TF1-30001. \textit{Bottom right}: SM1-TF1-30001.
		\label{fig:sgrams}}
\end{figure}

\subsection{Error analysis}
\label{ssec:error}

We demonstrate why our proposed method stayed comparable in terms of the naturalness with the two spectrogram illustrations shown in Figure~\ref{fig:sgrams}. Note that although the two illustrated samples come from conversion pairs that are not in the subjective evaluation set, these are indeed problems that exist in our proposed system since they are also occasionally found in the evaluation set. First, the wrapped frequency axes in the F0 transformed residual are very sensitive to imperfect converted spectral estimation. In our internal evaluation, we found that the husky characteristic of male speakers is hard to model due to the disadvantage condition of non-parallel training. Thus, the interaction of the imperfectly estimated converted feature and the F0 transformed residual signal tends to make the converted speech unstable. It could be observed in the bottom left of Figure~\ref{fig:sgrams} that the low frequency components tend to be very noisy, causing a seriously deteriorated voice. Informal comments from subjective test participants also confirmed this result.

Another problem worth noticing is that sometimes even by substituting features, the collapsed waveform problem is not completely avoided. As shown in the bottom right of Figure~\ref{fig:sgrams}, a short collapsed segment still exists even after the substitution process. This is a fundamental issue the DIFFVC is faced with when using a WSOLA plus resampling based F0 transformation process, since this problem exists in not only our proposed framework but also the original DIFFVC framework \cite{sprocket}. 

\section{Conclusions and Future Work}

In this paper, we introduced a generalization of the DIFFVC framework to make it applicable to general VC models. The proposed method is based on an F0 transformation in the residual domain, so that synthesis filtering is performed directly using the converted spectral features, thus removing the need for the conversion model to be able to predict the spectral differential, making the entire process free of parallel training data. We also introduced several techniques used in this framework, including 1) an alternative for high frequency component reconstruction based on zero stuffing, 2) collapsed waveform detection and corresponding feature substitution, and 3) power compensation due to resampling. Experimental results confirmed that when applied to a non-parallel VAE-based VC model, our method outperformed the counterpart that used a conventional vocoder in terms of conversion accuracy, yet the naturalness was on par. 

The investigation of the effectiveness of individual components proposed in this work, as well as a more in-depth analysis of the experimental results will be of top prior for future work. We also plan to apply our framework to other non-parallel VC models to further validate the effectiveness. Developing a frequency-wrapping robust spectral feature extractor may help solve the various issues discussed in Section~\ref{ssec:error}, which will be another important future work.

\noindent\textbf{Acknowledgements}: This work was partly supported by JST, PRESTO Grant Number JPMJPR1657 and JSPS KAKENHI Grant Number 17H01763, as well as the MOST-Taiwan Grants 107-2221-E-001-008-MY3 and 108-2634-F-001-004.

\bibliographystyle{IEEEtran}

\bibliography{is2019}

\end{document}